# A Scored Non-Deterministic Finite Automata Processor for Sequence Alignment

Ryan Karbowniczak          Rasha Karakchi

University of South Carolina
Karbowniczak@email.sc.edu
karakchi@cec.sc.edu

With the rapid expansion of symbolic data in fields such as internet data, biological data, and financial data, the need for efficient pattern matching and regular expression processing has surged [1–3, 8]. Some Pattern matching computations are modeled as Non-deterministic Finite Automata (NFA) [4, 5]. NFAs can have multiple states activated simultaneously, allowing concurrent operation of multiple next-state functions. This requires memory bandwidth that scales with state activation rates which often results in memory bottlenecks when they are executed on general-purpose platforms.

This inefficiency has driven interest in Domain-Specific Architectures (DSA) that can effectively utilize NFA parallelism. Many current automata processors are built using specialized hardware such as FPGAs and ASICs [6, 7], which enable efficient parallel processing. These specialized processors are faster than general-purpose CPUs for specific tasks and offer better energy efficiency and cost-effectiveness compared to traditional computing methods.

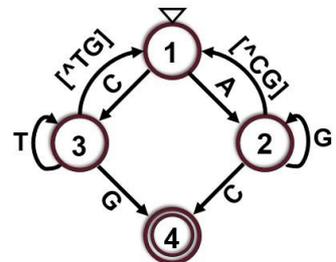

Recent research indicates that while existing automata processors are proficient at reporting pattern matches, many modern applications require identifying the best match path among multiple possible paths such as shortest path algorithms in graphs, probabilistic models, and sequence alignment in DNA sequences [1, 8]. For example, additive scoring function with gap penalty is used to measure the similarity between the aligned letters in Biology. Fig.1 shows an example of DNA dataset modeled as NFA. Given the match, mismatch and gap scores are +2, -1, -2 respectively, the highest score for sequence AGC is 6 and for sequence AGATG is -1.

The main objective of this work is to enable the existing FPGA-based automata processor to report the best sequence alignment represented by the highest score. Our approach is adopted from weight finite automata which requires incorporated weights (scores) into automata transitions to find the accumulated highest score and assess the output quality.

This approach, however, comes with several challenges that are addressed in this work. First, assigning values into transition introduces higher level of complexity because of the possible increase in state space complexity. Furthermore, due to the non-deterministic nature of the model, running multiple paths simultaneously can significantly increase the memory requirements and computational costs. For this reason, efficient implementation must balance performance and resource utilization. As the automaton expands, managing and processing weights becomes increasingly challenging and scalability is a major concern, particularly for large-scale systems.

In this work, we target NAPOLY, a Non-deterministic Finite Automata overlay which is a pattern-matching automata processing accelerator designed based on the alternative Micron-form of NFA description known as ANML (Automata Network Markup Language) and associates transition labels with states instead of edges [7]. Each state is called a State Transition Element (STE). The edges of NAPOLY, known as Fan-out, represent the point-to-point interconnections between STEs. We expanded NAPOLY by incorporating score values into the State Transition Elements (STE) and added arithmetic components to accumulate scores along the path and determine the final score. We refer to this extended STE as STE+ and to the proposed design as NAPOLY+.

If an STE+ is activated by a predecessor and the input symbol matches the transition symbol, the STE+ score is calculated based on the incoming score and the edge score, which is a fixed number stored in a register and configured during the reconfiguration stage. To handle the possibilities of mismatches and gaps, we designed all (STE+)s to be connected to the start STE+. However, this implementation is constrained by the array size and the maximum fan-out which is limited by the interconnection design and hardware resources. The interconnections are structured as a grid of global (horizontal) wires and local (vertical) wires on the fine-grained layer of the chip. The horizontal wires are limited by the maximum bus size that can be created on the selected FPGA (1 million wires), while the vertical wires represent the local fan-out wires connected to and from the STE+.

During the operation, start STE+ remains active at all times, enabling the processing of multiple symbols simultaneously and allowing multiple STE+ to be active at once. To distinguish between new symbol signals and mismatch incoming signals, we initialize the incoming scores to zero for every new symbol, indicating the possibility of a new path. We allocated a specific fan-in signal exclusively for the new symbol. The accepting (STE+)s are designed similarly to regular STE but have no connection with the start state since a match is found and score is reported once the accepting STE+ is activated. Both pattern sets and input sequences are stored in buffers, while the vector of accepting state IDs, input symbol offsets, and scores are flushed out to an output buffer. All buffers are assumed to be connected with DRAMs.

We evaluated our design on the Zynq Ultrascale+ ZCU104 FPGA Board, which includes 225K LUTs, 445K registers, and 64K distributed memory. Transition symbols were stored in Distributed Memory, while transition scores were stored in a register. We varied the array size from 1K to 64K and adjusted the maximum number of fan-outs. The results show that over 95% of device utilization (LUTs and registers) was achieved across all configurations. Distributed memory usage increased proportionally with the array size, whereas the maximum frequency (Fmax) decreased as the array size grew. Our results focus solely on the STE+ array, excluding buffers and DRAMs. Throughput was measured based on the number of input symbols (in Bytes) streamed and processed on NAPOLY+, without accounting for the time required to flush input buffers, matches, scores to the output buffer, or reconfiguration time. These results are preliminary, as our goal is to evaluate the end-to-end design using actual datasets, such as BLAST.

# 1 Acknowledgments

This research was supported by McNair Junior Fellowship at Molinaroli College of Engineering and Computing at USC. Author travel was partially supported by the USC Office of Undergraduate Research Magellan Voyager travel award and by ACM's Special Interest Group on High Performance Computing (SIGHPC).